\documentclass[aps,prl,preprint,superscriptaddress]{revtex4}
\usepackage{graphicx}
\usepackage{color}
\usepackage{amsfonts}
\usepackage{isomath}
\usepackage{amsmath}
\usepackage{amsthm}
\usepackage{amsfonts}
\usepackage{amssymb}
\usepackage{fdsymbol}
\usepackage{float}
\usepackage{braket}

\newcommand{\EF}{$E_\mathrm{F}$}

\def\kbar{$\bar{\mathrm{K}}$}
\def\mbar{$\bar{\mathrm{M}}$}

\def\gbar{$\bar{\mathrm{\Gamma}}$}

\def\WS2{WS$_2$}
\def\MoS2{MoS$_2$}

\def\EF{$E\mathrm{_F}$}

\renewcommand{\vec}[1]{\mathbfit{#1}}

\begin{document}

\title{Ultrafast triggering of insulator-metal transition in two-dimensional VSe$_2$}

\author{Deepnarayan Biswas}
\affiliation{Department of Physics and Astronomy, Interdisciplinary Nanoscience Center, Aarhus University,
8000 Aarhus C, Denmark}
\author{Alfred J. H. Jones}
\affiliation{Department of Physics and Astronomy, Interdisciplinary Nanoscience Center, Aarhus University,
8000 Aarhus C, Denmark}
\author{Paulina Majchrzak}
\affiliation{Department of Physics and Astronomy, Interdisciplinary Nanoscience Center, Aarhus University,
8000 Aarhus C, Denmark}
\affiliation{Central Laser Facility, STFC Rutherford Appleton Laboratory, Harwell 0X11 0QX, United Kingdom}
\author{Byoung Ki Choi}
\affiliation{Department of Physics, University of Seoul, Seoul 02504, Republic of Korea}
\author{Tsung-Han Lee}
\affiliation{Department of Physics and Astronomy, Rutgers University, Piscataway, New Jersey 08856, USA}
\author{Klara Volckaert}
\affiliation{Department of Physics and Astronomy, Interdisciplinary Nanoscience Center, Aarhus University,
8000 Aarhus C, Denmark}
\author{Jiagui Feng}
\affiliation{SUPA, School of Physics and Astronomy, University of St Andrews, St Andrews KY16 9SS, United Kingdom}
\affiliation{Suzhou  Institute  of  Nano-Tech.   and  Nanobionics  (SINANO),CAS,  398  Ruoshui  Road,  SEID,  SIP,  Suzhou,  215123,  China}
\author{Igor Markovi\'c}
\affiliation{SUPA, School of Physics and Astronomy, University of St Andrews, St Andrews KY16 9SS, United Kingdom}
\affiliation{Max Planck Institute for Chemical Physics of Solids, N\"othnitzer Stra{\ss}e 40, 01187 Dresden, Germany}
\author{Federico Andreatta}
\affiliation{Department of Physics and Astronomy, Interdisciplinary Nanoscience Center, Aarhus University,
8000 Aarhus C, Denmark}
\author{Chang-Jong Kang}
\affiliation{Department of Physics and Astronomy, Rutgers University, Piscataway, New Jersey 08856, USA}
\author{Hyuk Jin Kim}
\affiliation{Department of Physics, University of Seoul, Seoul 02504, Republic of Korea}
\author{In Hak Lee}
\affiliation{Department of Physics, University of Seoul, Seoul 02504, Republic of Korea}
\author{Chris Jozwiak}
\affiliation{Advanced Light Source, E. O. Lawrence Berkeley National Laboratory, Berkeley, California 94720, USA}
\author{Eli Rotenberg}
\affiliation{Advanced Light Source, E. O. Lawrence Berkeley National Laboratory, Berkeley, California 94720, USA}
\author{Aaron Bostwick}
\affiliation{Advanced Light Source, E. O. Lawrence Berkeley National Laboratory, Berkeley, California 94720, USA}
\author{Charlotte E. Sanders}
\affiliation{Central Laser Facility, STFC Rutherford Appleton Laboratory, Harwell 0X11 0QX, United Kingdom}
\author{Yu Zhang}
\affiliation{Central Laser Facility, STFC Rutherford Appleton Laboratory, Harwell 0X11 0QX, United Kingdom}
\author{Gabriel Karras}
\affiliation{Central Laser Facility, STFC Rutherford Appleton Laboratory, Harwell 0X11 0QX, United Kingdom}
\author{Richard T. Chapman}
\affiliation{Central Laser Facility, STFC Rutherford Appleton Laboratory, Harwell 0X11 0QX, United Kingdom}
\author{Adam S. Wyatt}
\affiliation{Central Laser Facility, STFC Rutherford Appleton Laboratory, Harwell 0X11 0QX, United Kingdom}
\author{Emma Springate}
\affiliation{Central Laser Facility, STFC Rutherford Appleton Laboratory, Harwell 0X11 0QX, United Kingdom}
\author{Jill A. Miwa}
\affiliation{Department of Physics and Astronomy, Interdisciplinary Nanoscience Center, Aarhus University,
8000 Aarhus C, Denmark}
\author{Philip Hofmann}
\affiliation{Department of Physics and Astronomy, Interdisciplinary Nanoscience Center, Aarhus University,
8000 Aarhus C, Denmark}
\author{Phil D. C. King}
\affiliation{SUPA, School of Physics and Astronomy, University of St Andrews, St Andrews KY16 9SS, United Kingdom}
\author{Young Jun Chang}
\affiliation{Department of Physics, University of Seoul, Seoul 02504, Republic of Korea}
\author{Nicola Lanata}
\affiliation{Department of Physics and Astronomy, Interdisciplinary Nanoscience Center, Aarhus University,
8000 Aarhus C, Denmark}
\author{S{\o}ren~Ulstrup}
\affiliation{Department of Physics and Astronomy, Interdisciplinary Nanoscience Center, Aarhus University,
8000 Aarhus C, Denmark}
\email[Electronic address: ]{ulstrup@phys.au.dk}  

\maketitle

\textbf{Assembling transition metal dichalcogenides (TMDCs) at the two-dimensional (2D) limit is a promising approach for tailoring emerging states of matter such as superconductivity or charge density waves (CDWs)  \cite{Xi2015, Ugeda2016, Sanders2016, Pasztor2017, Chen:2020}. Single-layer (SL) VSe$_2$ stands out in this regard because it exhibits a strongly enhanced CDW transition with a higher transition temperature compared to the bulk in addition to an insulating phase with an anisotropic gap at the Fermi level \cite{Ganbat:2018,Feng:2018, Chen:2018,Strocov:2012,Umemoto2019}, causing a suppression of anticipated 2D ferromagnetism in the material \cite{Feng:2018,Bonilla2018,Coelho:2019,Wong2019}. Here, we investigate the interplay of electronic and lattice degrees of freedom that underpin these electronic phases in SL VSe$_2$ using ultrafast pump-probe photoemission spectroscopy. In the insulating state, we observe a light-induced closure of the energy gap on a timescale of 480 fs, which we disentangle from the ensuing hot carrier dynamics. Our work thereby reveals that the phase transition in SL VSe$_2$ is driven by electron-lattice coupling and demonstrates the potential for controlling electronic phases in 2D materials with light.}

Switching between a normal and an unconventional phase of a material using an ultrafast laser pulse provides an opportunity to probe fundamental interactions and determine how they concur in driving the phase transition \cite{Cavalleri:2001}. This procedure has been employed in time- and angle-resolved photoemission spectroscopy (TR-ARPES) experiments to observe energy-, momentum- and time-dependent melting of CDW and Mott insulator phases in bulk TMDCs \cite{Perfetti:2006,Hellmann:2012,Rohwer:2011, Petersen:2011,Mathias:2016} and uncover Cooper pair recombination rates in high-temperature superconductors \cite{Graf:2011,Smallwood1137}. The timescales on which the electronic system evolves following excitation provide detailed insights into the hierarchy of interactions underpinning the phase transition. For example, electronic degrees of freedom typically respond on timescales in the range of $10 -100$~fs \cite{Hellmann:2012,Sohrt2014}, whereas processes involving lattice degrees of freedom occur on timescales longer than 100~fs \cite{Perfetti:2007,Monney2016}.  

\begin{figure*} [t!]
\begin{center}
\includegraphics[width=1\textwidth]{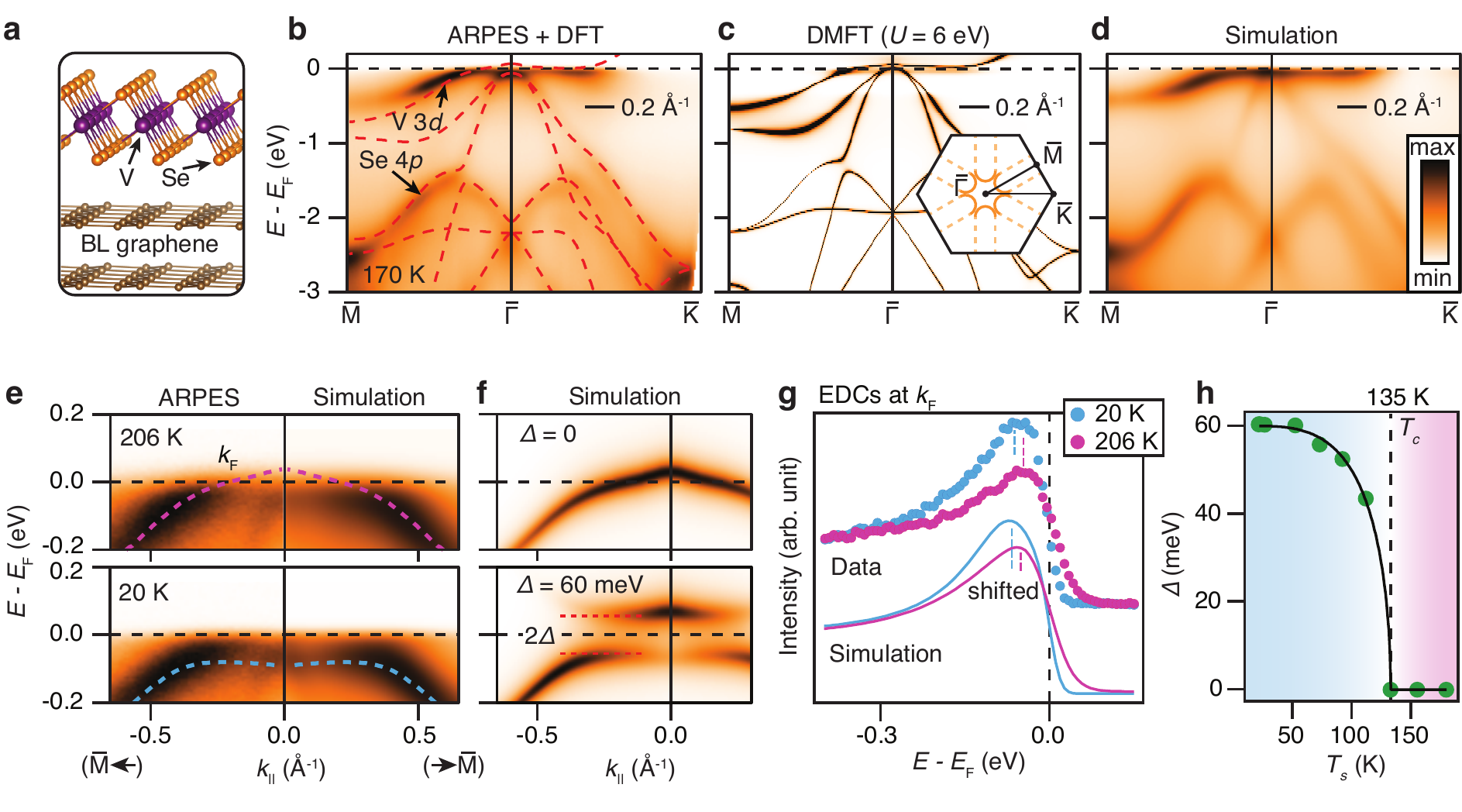}
\caption{\textbf{Electronic structure and temperature dependent phase transition in SL VSe$_2$.} \textbf{a,} Schematic of 1T structure of SL VSe$_2$. \textbf{b,} ARPES data collected at 170 K in the \mbar-\gbar-\kbar~direction. The dashed red lines correspond to the calculated DFT dispersion (the raw data without DFT bands is shown in Supplementary Fig. S3). \textbf{c,} DMFT spectrum for $U=6$~eV. The inset presents the SL VSe$_2$ BZ with contours sketching the Fermi surface at 170~K. \textbf{d,} Numerical simulation of the 2D ARPES intensity optimized to the data in (\textbf{b}).  \textbf{e,} ARPES spectra and corresponding simulation along \mbar-\gbar~for the given sample temperatures (see Supplementary Video 1 and Supplementary Fig. S3). The dashed lines represent the simulated dispersion including many-body effects. \textbf{f,} Simulation of photoemission intensity with the Fermi-Dirac function removed for the given values of the gap parameter $\Delta$. \textbf{g,} Comparison of EDCs at $k_{\mathrm F}$ for sample temperatures of 206 and 20~K. The cuts were obtained from the data (markers) and simulation (curves) in \textbf{e}. Tick marks indicate EDC peak positions, demonstrating a shift away from \EF~with decreasing sample temperature. \textbf{h,} Extracted values (markers) of $\Delta$ from spectra obtained at several sample temperatures along with a fit (black curve) to a mean-field expression describing the phase transition. The fitted value of $T_c$ is shown via a dashed vertical line.}
\label{fig:1}
\end{center}
\end{figure*}

We apply TR-ARPES in order to investigate the microscopic origin of the insulator-metal transition in SL VSe$_2$ grown on bilayer (BL) graphene \cite{Ganbat:2018,Feng:2018}. Before discussing the results of pump-probe experiments, we start by clarifying the electronic structure of the material and characterize the phase transition in static conditions. Above the critical temperature $T_c$, the material assumes the 1T structural modification where the V and Se atoms are coordinated in an octahedral geometry as shown in Fig. \ref{fig:1}(a). The dispersion of this phase is calculated using density functional theory (DFT) and presented together with an ARPES spectrum along the high symmetry \mbar-\gbar-\kbar~direction in Fig. \ref{fig:1}(b). The shallow states around \EF ~are composed of V 3$d$ orbitals while the dispersive subbands at higher binding energies derive from Se 4$p$ orbitals (see Supplementary Fig. S1). The increased broadening of V 3$d$ states with energy below \EF~and the enhanced effective masses compared to the DFT dispersion are the telltale signs of correlation effects. Indeed, we find that these effects are captured in our dynamical mean-field theory (DMFT) calculations for a relatively large Hubbard interaction strength $U$ as shown in Fig. \ref{fig:1}(c) for  $U = 6$~eV (see also Supplementary Fig. S2). 

In order to deconvolve single-particle and many-body effects from the ARPES spectra we are following the phenomenological model of the photoemission intensity, ${\cal I}(k, \omega) = |{\cal M}(k,\omega)|^2 {\cal A}(k, \omega) n_{\mathrm{FD}}(\omega)$ \cite{Damascelli:2003}. Here, $\mathcal{A}(k, \omega)$ is the spectral function, ${\cal M}(k, \omega)$ incorporates the single-electron dipole matrix elements that govern the selection rules of the photoemission process and $n_{\mathrm{FD}} =(e^{(\omega-\mu)/k_{\mathrm B}T_{e}}+1)^{-1}$ is the Fermi-Dirac (FD) distribution function with chemical potential $\mu$ and electronic temperature $T_e$. By combining the bare dispersion obtained from DFT with the electronic self-energy, $\Sigma$, deduced from the DFT and LDA + DMFT calculation, we are able to construct the spectral function of SL VSe$_2$, as described in further details in the Methods. A numerical simulation of ${\cal I}(k, \omega)$ with self-energy and matrix elements adjusted to give an optimum description of the measured 2D image of the photoemission intensity is shown in Fig. \ref{fig:1}(d), providing a basic model to interpret ARPES spectra of SL VSe$_2$ in the following discussion.

\begin{figure*}[t!]
\begin{center}
\includegraphics[width=1\textwidth]{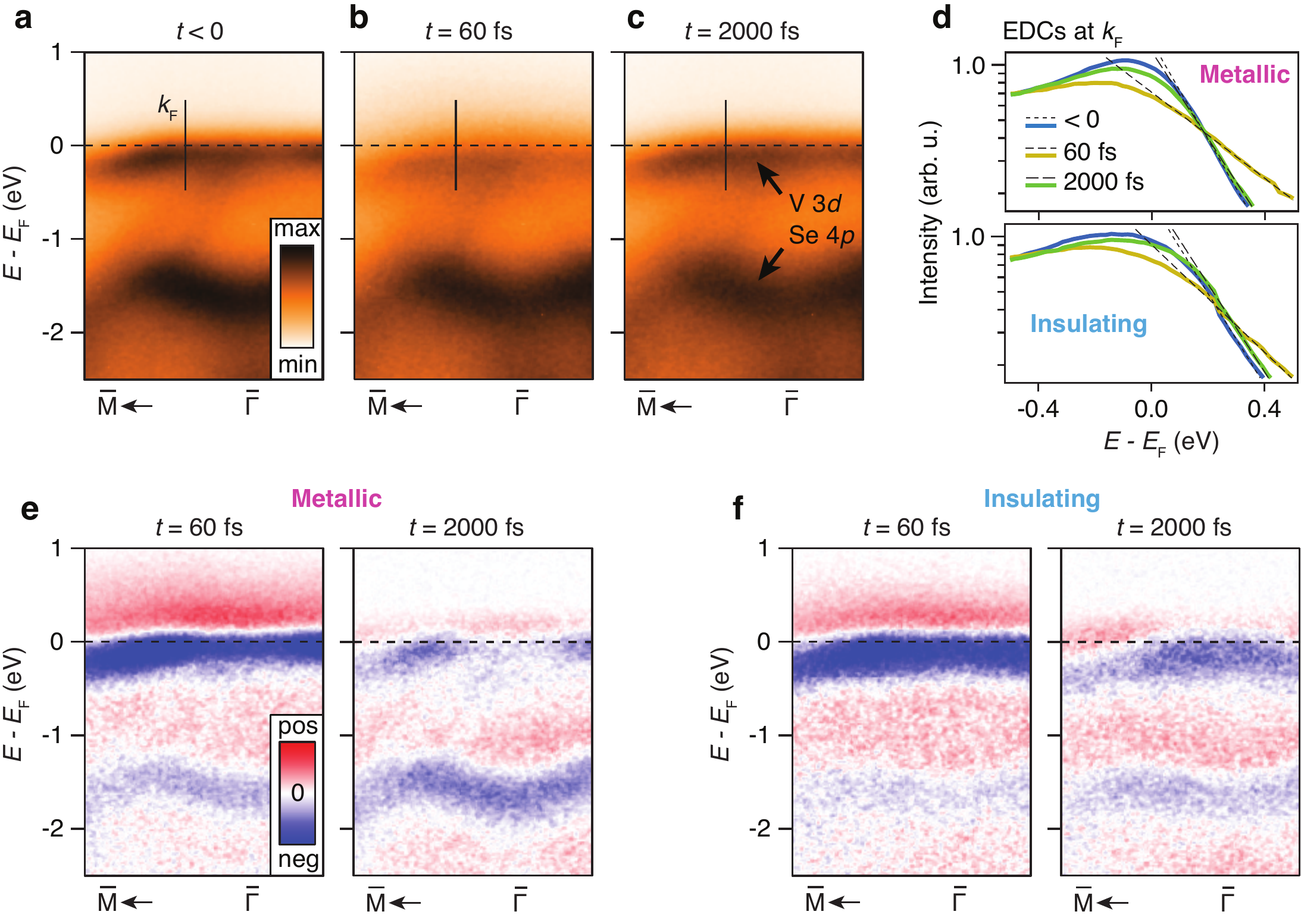}
\caption{\textbf{Optical excitation of SL VSe$_2$.} \textbf{a}-\textbf{c,} ARPES spectra of V 3$d$ and Se 4$p$ bands (see arrows in \textbf{c}) revealing the response of the electronic structure \textbf{a} before ($t<0$), \textbf{b} at the peak ($t=60$~fs) and \textbf{c} at a long delay ($t=2000$~fs) after optical excitation. All spectra are obtained along \mbar-\gbar~direction and with the sample temperature initially at 200~K. \textbf{d,} EDCs extracted along the vertical line at $k_\mathrm{F}$ shown in \textbf{a}-\textbf{c} for the corresponding time delays. The top panel displays data for the metallic phase ($T_s = 200$~K) while the bottom panel presents EDCs for the insulating phase ($T_s = 88$~K). The intensity is plotted on a logarithmic scale. The dashed lines are exponential function fits to the tail of the EDCs. \textbf{e,} Intensity difference for the metallic phase obtained by subtracting the equilibrium spectrum in \textbf{a} from the excited state spectra at the given time delays in \textbf{b}-\textbf{c}. \textbf{f,} Similar difference spectra as shown in \textbf{e} measured for the insulating phase.} 
\label{fig:2}
\end{center}
\end{figure*}

We focus on measurements along the \mbar-\gbar~high symmetry direction in order to track the opening of a gap in the Fermi surface segment shown in the Brillouin zone (BZ) sketch in Fig.  \ref{fig:1}(c), which occurs when the sample is cooled below $T_c$~\cite{Ganbat:2018,Umemoto2019}. A detailed view of this cut is presented for sample temperatures $T_s$ of 206 and 20~K around the Fermi crossing $k_\mathrm{F}$ in Fig.  \ref{fig:1}(e). A significant $T_s$-dependent change of the V 3$d$ dispersion is seen via the purple and blue dashed curves that have been extracted using our 2D simulation of the intensity (see transition in Supplementary Video 1). The change of dispersion is linked to the gap opening described in terms of the parameter $\Delta$, which is demonstrated in the simulation with the FD function removed in Fig.  \ref{fig:1}(f). For $\Delta = 0$ the band crosses \EF, however, the intensity around the band maximum is not seen in the ARPES data because these states are unoccupied. As $\Delta$ assumes a finite value, a gap of $2\Delta$ opens, leading to increased spectral weight around \gbar~below \EF. The presence of such a gap is further corroborated by a shift of the peak away from \EF~in energy distribution curve (EDC) cuts at $k_\mathrm{F}$ as $T_s$ is lowered, which is demonstrated in Fig. \ref{fig:1}(g). The complete $T_s$-dependence of $\Delta$ is determined by fitting the 2D ARPES intensity measured at several temperatures, leading to the phase diagram in Fig. \ref{fig:1}(h). The critical temperature found using this method is $135\pm5$ K, which is consistent with previous studies \cite{Feng:2018,Ganbat:2018}.

On the basis of the spectroscopic signatures and modeling of the insulator-metal transition specified above, we are now able to analyze the time-dependent response of SL VSe$_2$ to an optical excitation. Measurements performed with sample temperatures of 200~K and 88~K are compared in order to track the dynamics in both the metallic and insulating phases. TR-ARPES snapshots are shown along \mbar-\gbar~in Figs. \ref{fig:2}(a)-(c) for excitation of the metallic phase using a pump pulse with an energy of 1.56 eV, temporal width of 30~fs and a fluence around 5 mJ/cm$^{2}$ at a time delay $t$ before the optical excitation ($t<0$), at the peak of the excitation ($t=60$~fs) and at a longer delay ($t=2000$~fs). The excitation leads to a substantial decrease in intensity in the V 3$d$ states around \EF~(see panel~(b)), which does not fully recover at longer delays (see panel~(c)). The raw $(\omega,k,t)$-dependent intensity measured under the same conditions for the insulating phase appears similar on a superficial view (see Supplementary Fig. S3 for a comparison). A comparison of EDCs at $k_\mathrm{F}$, shown on a logarithmic intensity scale in Fig. \ref{fig:2}(d), reveals exponential tails with a $t$-dependent slope indicating the generation of hot carriers with an elevated electronic temperature in both phases.

A stronger indication for the spectral changes following excitation is obtained by calculating the difference in photoemission intensity by subtracting a spectrum determined in equilibrium conditions before the arrival of the pump pulse from the spectrum measured at a given delay time as shown for the two phases in Figs. \ref{fig:2}(e)-(f) and Supplementary Video 2. A highly complex $\omega$- and $k$-dependence of intensity depletion and increase is seen. Surprisingly, we observe strong difference signals persisting at long delays ($t=2000$~fs) that look dramatically different for the two phases. Naively, one could think of assigning these changes to the mere redistribution of charge carriers in the V 3$d$ and Se 4$p$ states with excited holes (electrons) signified by the blue (red) regions of the spectra. However, the intensity is simultaneously affected by a change of the FD distribution due to the elevated electronic temperature $T_e$, a $t$-dependence of the quasiparticle scattering rate $\Gamma$ that manifests itself as increased broadening of the bands \cite{Andreatta:2019,Ulstrup_2015}, and the possibility of a $t$-dependent $\Delta$ in the insulating phase. Using our model of the photoemission intensity presented in Fig. \ref{fig:1}(d) we fit $T_e$, $\Gamma$ and $\Delta$ such that our simulated intensity gives an optimum description of the ARPES intensity at all measured time delays, noting that $\Delta=0$ for the metallic phase (see Methods and Supplementary Section 3 for further details of the fit). An excellent fit is obtained for all time delays in both phases using our assumption of a hot carrier model where the simulated intensity always incorporates a well-defined FD function, indicating that thermalization occurs via electron-electron interactions on a faster timescale than we can resolve ($<40$ fs) \cite{Johannsen:2013ab}.  

\begin{figure*}[t!]
\begin{center}
\includegraphics[width=1\textwidth]{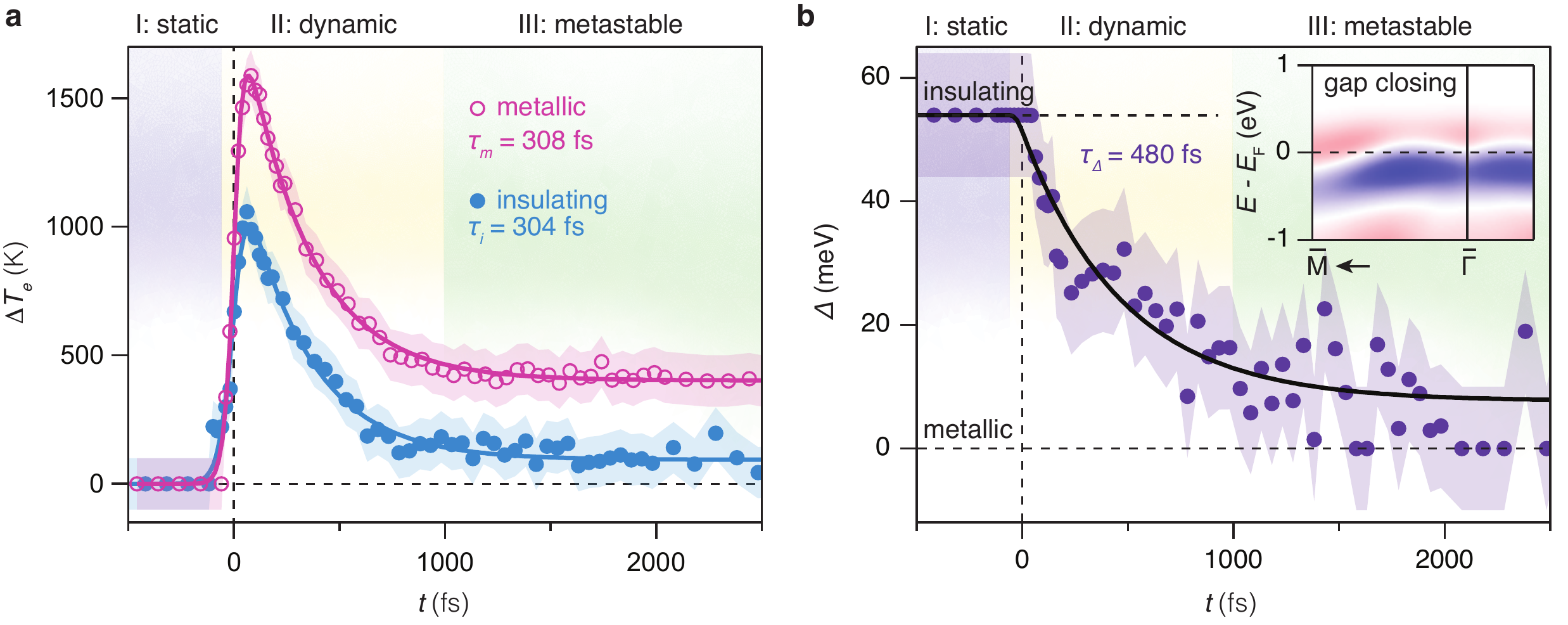}
\caption{\textbf{Dynamics of hot carriers and phase transition.}  \textbf{a,} Time-dependent change of electronic temperature determined by fitting the full $(\omega,k)$-dependence of the photoemission intensity at each measured time delay in the metallic (open purple circles) and insulating (filled blue circles) phases. The full curves are fits to a function consisting of an exponential rise followed by an exponential decay with the given time constants $\tau_{\mathrm{m(i)}}$ for the decay in the metallic (insulating) phase. The shaded areas correspond to three distinct periods of the time-dependent response labeled as (I) static, (II) dynamic and (III) metastable. \textbf{b,} Time-dependence of $\Delta$. The solid curve is a fit to an exponential decay with a time constant of $\tau_{\Delta} = 480$~fs. The inset presents the intensity difference between a fitted spectrum for the insulating state at equilibrium and a spectrum for the metastable metallic state, incorporating a change of $\Delta$ from 54~meV to zero. The shaded regions around the data points in \textbf{a}-\textbf{b} represent the uncertainity associated with the analysis.}
\label{fig:3}
\end{center}
\end{figure*}

The change of electronic temperature extracted from this analysis is presented in Fig. \ref{fig:3}(a), revealing a qualitatively similar $t$-dependence in the two phases that can be divided according to (I) a static period before excitation, (II) a dynamic period with a sharp rise of $T_e$ during excitation followed by an initial fast relaxation and (III) a metastable period where the system remains out of equilibrium and does not relax on the timescale of our measurement. The transient increase in electronic temperature is caused by ultrafast energy transfer from the laser pulse to the electrons in the V 3$d$ states. Energy is then transferred from electrons to the lattice leading to a decay of $T_e$ on a timescale of $\approx$300~fs. Both insulating and metallic phases subsequently reach a stable elevated electronic temperature compared to equilibrium, indicating that electron and lattice subsystems have reached a thermal equilibrium with a temperature larger than $T_c$.

Figure \ref{fig:3}(b) presents the extracted $t$-dependence of $\Delta$ from the TR-ARPES measurement of the insulating phase, revealing a transient closure of the gap during the dynamic period (region II), leaving the system in a metastable metallic phase (region III). The intensity difference between the fitted spectrum in this metastable metallic state and the insulating state in equilibrium (region I) essentially reproduces the measured behavior, as seen by comparing the inset in Fig. \ref{fig:3}(b) with the difference spectrum at $t=2000$ fs in Fig. \ref{fig:2}(f) (see Supplementary Fig. S5 for other fit parameters and fitted difference spectra in the metallic phase). Note that this distinct redistribution of intensity is highly reproducible over multiple samples as shown in Supplementary Fig. S6  and the change in $\Delta$ is essential for its simulation. The timescale where $\Delta$ goes to zero is comparable to the time it takes the electronic system to transfer energy to the lattice as seen in region II in Fig. \ref{fig:3}(a). As the lattice is thermally excited it obtains sufficient energy to rearrange atoms and trigger the insulator-metal transition, which is clocked to the time constant $\tau_{\Delta} = 480$~fs obtained from an exponential fit. This is on the order of quenching times observed for strong electron-lattice coupling CDWs in bulk TMDCs and significantly slower than the timescales associated with melting of Mott gaps driven by electron-electron interactions \cite{Sohrt2014}. 

The appearance of a metastable state is strongly indicative of a slow reconfiguration of the thermally excited lattice, possibly involving the distorted ($\sqrt{3} \times 2$) and ($\sqrt{3} \times \sqrt{7}$) superstructures found in the insulating phase of SL VSe$_2$ \cite{Feng:2018,Ganbat:2018,Chen:2018,Wong2019}. This situation bears a striking resemblance to VO$_2$ where an ultrafast excitation transforms an insulating phase with a monoclinic structure to a metallic phase with a rutile structure \cite{Cavalleri:2001,Cavalleri:2004,Kubler:2007}. Such dynamics in SL VSe$_2$ may be resolved in future studies utilizing ultrafast probes of the lattice structure. 

In conclusion, we have tracked the spectral function of SL VSe$_2$ across an ultrafast insulator-metal transition triggered by an intense optical excitation. The spectroscopic signatures of hot carrier dynamics and phase transition could be disentangled, revealing that electron-lattice energy exchange drives the transition in the first few hundreds of femtoseconds following excitation and leads to a metastable metallic state. Such a situation is not only intriguing for the application of 2D materials in electronic memory devices, but the coupling between electron and lattice degrees of freedom is also of fundamental interest for understanding the interplay of CDW physics and magnetism in 2D.

\section{Acknowledgement}
We thank Phil Rice, Alistair Cox and David Rose for technical support during the Artemis beamtime. We gratefully acknowledge funding from VILLUM FONDEN through the Young Investigator Program (Grant. No. 15375) and the Centre of Excellence for Dirac Materials (Grant. No. 11744), the Danish Council for Independent Research, Natural Sciences under the Sapere Aude program (Grant Nos.  DFF-9064-00057B and DFF-6108-00409) and the Aarhus University Research Foundation. This work is also supported by National Research Foundation (NRF) grants funded by the Korean government (nos. NRF-2020R1A2C200373211 and 2019K1A3A7A09033389) and by the International Max Planck Research School for Chemistry and Physics of Quantum Materials (IMPRS-CPQM). The authors also acknowledge The Royal Society and The Leverhulme Trust. Access to the Artemis Facility was funded by STFC. The Advanced Light Source is supported by the Director, Office of Science, Office of Basic Energy Sciences, of the U.S. Department of Energy under Contract No. DE-AC02-05CH11231.

\section{Competing Interests}
The authors declare that they have no competing financial interests.

\section{Additional information}

\textbf{Supplementary Information} and two Supplementary Videos accompany this paper.\\

\textbf{Correspondence and requests for materials} should be addressed to S. U. (ulstrup@phys.au.dk).

\section{Methods}
\textbf{Sample preparation.} SL VSe$_2$ samples were grown on BL graphene on 6H-SiC(0001) using molecular beam epitaxy (MBE) with a base pressures better than $2 \cdot 10^{-10}$~Torr. The sample measured to produce the data in Fig. \ref{fig:1}(b) was grown at the University of St Andrews, UK. The remaining spectra were collected on samples  grown at the University of Seoul, Republic of Korea. To obtain the BL graphene on SiC, the SiC substrates were outgassed at 650~$^{\circ}$C for a few hours and then annealed three times up to 1300~$^{\circ}$C for 2~min. The formation of BL graphene was verified by reflection high-energy electron diffraction (RHEED) and low-energy electron diffraction (LEED). High-purity V (99.8\%) and Se (99.999\%) were simultaneously evaporated while the substrate was kept at 250~$^{\circ}$C. The growth process was monitored with in situ RHEED.  The growth rate was fixed at 5~min per Se-V-Se layer. After growth, the sample was annealed at 450~$^{\circ}$C for 30 min. A 100~nm Se film was deposited at room temperature to protect the sample while transferring through air. This Se capping layer was removed by annealing the sample at 300~$^{\circ}$C for several hours in the UHV analysis chamber before photoemission experiments. No noticeable change in sample quality was observed due to the capping and de-capping procedure.\\

\textbf{Static ARPES experiments.} The ARPES spectrum shown in Fig. \ref{fig:1}(b) was collected using a high-intensity He lamp ($h\nu$ = 21.2~eV, p-polarization) and a SPECS Phoibos 225 hemispherical electron analyzer at University of St Andrews, UK. Here, samples were directly transferred from the attached MBE growth chamber to the ARPES chamber. The remaining static measurements were performed in the microARPES end-station (base pressure of $\sim$3~$\cdot~10^{-11}$~Torr) at the MAESTRO facility at beamline 7.0.2 of the Advanced Light Source, Lawrence Berkeley National Laboratory. The ARPES system was equipped with a Scienta R4000 electron analyzer. We used a photon energy of 48 eV for $T_s$-dependent scans. The total energy and angular resolution of our experiments were better than 20 meV and  $0.1^{\circ}$, respectively.\\ 

\textbf{TR-ARPES experiments.} The Materials Science end station of the Artemis facility at Rutherford Appleton Laboratory was used for TR-ARPES measurements. Synchronized infrared (IR) pump and extreme ultraviolet (EUV) probe beams were generated from a Ti:Sapphire laser system at 12 mJ, 1 kHz, with a 30 fs pulse length and a central wavelength of 795 nm. The output of the laser was split:  a small fraction of the energy was used directly to pump the sample at 1.56 eV, with fine control of the fluence achieved using a half waveplate and a thin film polariser, while 2 mJ pulse energy was used to prepare the probe beam by high-harmonic generation. The laser was focused into a thin Ar gas cell to generate a comb of odd harmonics in the EUV. The 19th harmonic at 29.6 eV was selected, using a time-preserving monochromator \cite{Fabio2011}. The two beam polarisations were orthogonal:  the pump beam was s-polarised while the probe beam was p-polarised. The end station is equipped with a SPECS Phoibos 100 hemispherical electron energy analyser. Experiments were performed in the wide-angle mode, with a slit size of 1 mm. The time and angular resolution of the experiments were 46 fs and 0.3$^{\circ}$, respectively.  The optimum energy resolution was 250 meV, as determined through our simulations of the photoemission intensity.  Energy resolution is limited by analyzer energy resolution (about 190 meV), EUV probe pulse broadening (about 100 meV), and space charge effects (about 130 meV). Temperatures from 88 to 220 K were reached using an open-cycle liquid He cryostat.\\

\textbf{Theory.} The LDA+DMFT calculations were performed at a temperature of $T_s= 200$~K, assuming a Hund's rule coupling $J=0.8$ eV and scanning different values of the screened Hubbard interaction strength $U$ from 5 to 9~eV. We utilized the DMFT package for the electronic structure calculations of Ref. \cite{PRB_Haule_2010} interfaced with the local density approximation (LDA) functional implemented in Wien2k \cite{wien2k} and we adopted the fully-localized-limit scheme for the double-counting functional \cite{JPCM_Anisimov_1997}. All simulations have been performed with 10000 $k$-points and $Rkmax=7$ and employing the continuous-time quantum Monte-Carlo (CTQMC) impurity solver \cite{RMP_Gull_2011, PRL_Werner_2006, PRB_Haule_2007}. The spectral properties were mapped onto the real axis using the maximum entropy method \cite{PR_Jarrell_1996}. The DFT calculations were performed using a Perdew, Burke, and Ernzerhof (PBE) functional \cite{PBE}.\\

\textbf{Simulation of ARPES intensity.} The photoemission intensity is described by the expression
\begin{align}
{\cal I}(\vec{k}, \omega) = |{\cal M}_n(\vec{k}, \omega)|^2 {\cal A}_n(\vec{k}, \omega) n_{\mathrm FD}(\omega,T_e),
\label{eq:transphoto1}
\end{align}
as explained in the main text. The subscript $n$ is the band index. The spectral function is written as \cite{Damascelli:2003, Nechaev2009, Andreatta:2019},
\begin{align}
{\cal A}_n(\vec{k}, \omega) = -\frac{\pi^{-1}{{\Sigma}^{\prime\prime}}_n(\vec{k}, \omega) }{\left(\omega-\epsilon_{\vec{k}n}- {{\Sigma}^{\prime}}_n(\vec{k}, \omega) \right)^2 +{{\Sigma}^{\prime\prime}}_n^2(\vec{k}, \omega) },
\label{eq:transphoto2}
\end{align}
where $\epsilon_{\vec{k}n}$ is the non-interacting band dispersion or bare band, which we describe using the DFT dispersion as an input. ${\Sigma}^{\prime}_n$ and ${\Sigma}^{\prime\prime}_n$ are the real and imaginary parts of the electronic self energy ${\Sigma}_n$, respectively. In our simulation the correlation effects in the V 3$d$ states, including the gap opening of the dispersion, are included in ${\Sigma}_n$ through the expression (see Supplementary Section 1 for further details)
\begin{align}
{{\Sigma}}_n({\vec{k}},\omega) = {\Sigma}_0 - \frac{1-Z}{Z} \omega  - i\frac{{\Gamma}}{Z}  
+ \frac{\Delta^2/Z}{\omega+Z(\epsilon_{\vec{k}n}+{\Sigma}_0)+ i{\Gamma}_0}.
\label{eq:transphoto3}
\end{align}
Here ${\Sigma}_0$ is a constant energy shift of the states, $Z$ is the quasiparticle residue, ${\Gamma}$ is quasi-particle scattering rate, $\Delta$ is the gap parameter and ${\Gamma}_0$ is a constant related to the change in the scattering rate due to the presence of a gap. We also used a parameter, labeled ${\mathrm\Delta}E_s$, to describe any rigid shift of all the bands including \EF. Such a shift may arise due to an external electric field associated with vacuum space charge or surface photo voltage \cite{ulstrup2015}. Finally, the ARPES intensity is obtained by convoluting the photoemission intensity ${\cal I}(\vec{k}, \omega) $ by two Gaussians representing the energy (${\cal R}_{\omega}$) and momentum ($ {\cal R}_k$) resolution broadening of the instrument,
\begin{align}
{\cal I}_{ARPES}={\cal I}(\vec{k}, \omega) \ast {\cal R}_{\omega}\ast {\cal R}_k.
\label{eq:transphoto4}
\end{align}

We expand ${\cal M}_n({\vec{k}},\omega)$ in second order polynomial terms of both $\omega$ and $k$ \cite{Andreatta:2019}. The Se\nolinebreak~4$p$ states at higher binding energies are well-described by the DFT bands and using a scattering rate that is merely expressed in terms of first order polynomials of  $\omega$ and $k$. The parameters describing ${\cal M}_n({\vec{k}},\omega)$, ${\Sigma}_n$ and $n_{FD}$ are found in static conditions by performing a 2D fit of the simulated intensity to the ARPES spectra. We find that a satisfactory fit is obtained using a quasiparticle residue $Z$ in the range of 0.52 to 0.54. For the fits of the TR-ARPES data we account for the time dependent changes of FD distribution and spectral function by allowing a variation of $T_e$, ${\Delta}E_s$ and the self-energy through the scattering rate ($\Gamma$) and the gap parameter $\Delta$. We allow for a slight adjustment of the energy- and momentum-position of the bands to ensure consistency between measured and fitted spectra. The resulting parameters of the fit to the TR-ARPES data are given in Fig. \ref{fig:3} and Supplementary Fig. S5.

\newpage

\subsection{Supplementary Section 1: Form of the self energy in ARPES simulations}

\begin{figure*}[ht]
\begin{center}
\includegraphics[width=.95\textwidth]{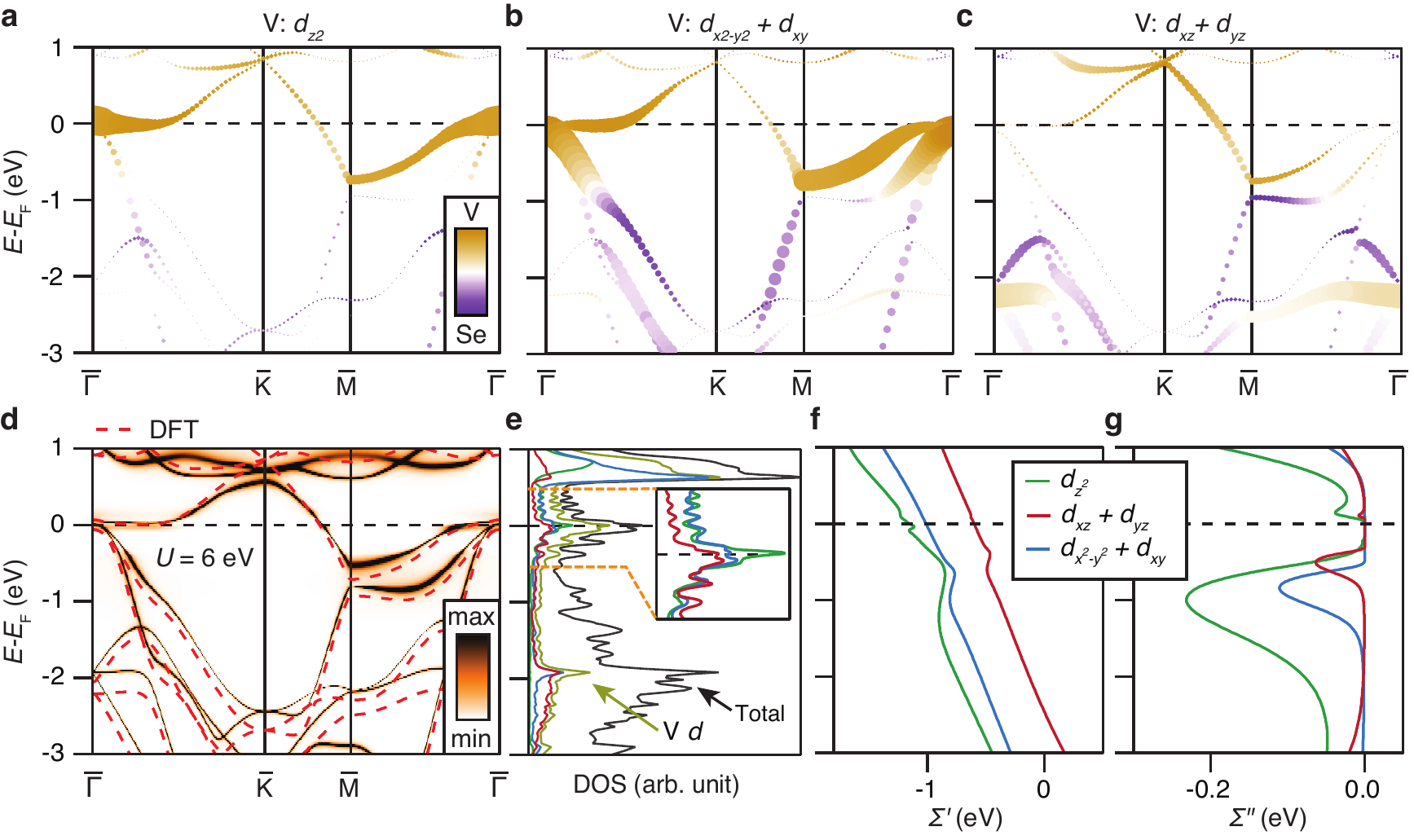}
\caption{\textbf{Theoretical calculations and band character.} \textbf{a}-\textbf{c,} DFT calculated bands. The contribution from V and Se atoms are represented by yellow and violet color.  The marker size corresponds to the V $d_{z^2}$, $d_{x^2+y^2}+d_{xy}$ and $d_{xz}+d_{yz}$ orbital characters.  \textbf{d}, LDA + DMFT calculated band structure ($U=$ 6 eV, $J=0.8$ eV). The DFT bands are overlayed as red dashed lines. \textbf{e}, Density of states corresponding to LDA + DMFT calculated band structure, shown in \textbf{(d)}. \textbf{f}-\textbf{g,} Energy dependence of the \textbf{f} real and \textbf{g} imaginary parts of the electronic self energy, respectively.}
\label{Supp:1}
\end{center}
\end{figure*}

The following mathematical form of the self energy was utilized in our simulations for interpreting the ARPES data, within an energy window of $\sim$0.5~eV from the Fermi level:
\begin{equation}
\label{eq:transphoto}
{\Sigma}_n({\vec{k}},\omega) = \Sigma_{loc}(\omega)  - i\frac{\Gamma}{Z} 
+ \frac{\Delta^2/Z}{\omega+Z(\epsilon_{\vec{k}n}+\Sigma_0)+ i\Gamma_0}\\
\,.
\end{equation}
Here $\vec{k}$ is the momentum, $\omega$ is the energy,
$\epsilon_{\vec{k}n}$ is a generic band eigenvalue,
$\Gamma$ is scattering rate, $\Delta$ is the gap parameter, $\Gamma_0$ is a constant related to the change in scattering rate due to $\Delta$ and $\Sigma_{loc}$ is the momentum-independent (local) component of the self-energy:
\begin{equation}
\Sigma_{loc}(\omega) = \Sigma_0 - \frac{1-Z}{Z} \omega 
\,,
\label{sloc}
\end{equation}
which we approximated assuming a linear structure characterized by
the quasiparticle residue $Z$ and a constant energy shift $\Sigma_0$.

To explain the physical reasons underlying  Eq.~\eqref{eq:transphoto} we note that the corresponding momentum-resolved single-particle Green's function is represented as follows:
\begin{equation}
G_n({\vec{k}},\omega)=\frac{1}{\omega-\epsilon_{\vec{k}n}-{\Sigma}_n(\omega,{\vec{k}})}=
\frac{Z}{\omega-\epsilon^*_{\vec{k}n}+\Gamma
-\frac{\Delta^{2}}{\omega+\epsilon_{\vec{k}n}+i\Gamma_0}}\,,
\label{qp}
\end{equation}
where
\begin{equation}
\epsilon^*_{\vec{k}n}=Z\, ( \epsilon_{\vec{k}n} + \Sigma_0 )\,.
\end{equation}
In fact, the last expression in Eq.~\eqref{qp} has the same mathematical structure of the phenomenological self-energy  previously used for fitting ARPES data in the presence of a superconducting  or charge density wave (CDW) gap \cite{BCSSE_Norman_1998, Chen:2018}.
Therefore, in this work, \emph{Eq.~\eqref{eq:transphoto} is designed to represent the CDW effects on a band structure consisting of pre-existing quasi-particle excitations renormalized by electron correlations.}

\begin {table}[t!]
\caption{Quasiparticle weights of V 3$d$ orbitals for different values of $U$, at $J=0.8$~eV}\label{table:z}
\smallskip
\centering
\begin{tabular}{| c | c | c | c |}
\hline
\textbf{$U$} & \textbf{$d_{z^2}$} & \textbf{$d_{x^2+y^2}+d_{xy}$} & \textbf{$d_{xz}+d_{yz}$} \\
\hline
5 eV~          & ~~0.50635~~      & ~~0.78723~~       & ~~0.80550~~        \\
6 eV~         & 0.49489      & 0.75884           & 0.77807        \\
7 eV~           & 0.47736      & 0.73362                  & 0.75279        \\
8 eV~           & 0.45801      & 0.70974            & 0.72878        \\
9 eV~           & 0.43984      & 0.68810                  & 0.70745       \\
\hline
\end{tabular}
\end{table}

Note that in Eq.~\eqref{sloc} we assumed that $\Sigma_{loc}(\omega)$ acts as a number rather than a matrix.
However, 
in general, the self-energy correction $\Sigma_{loc}(\omega)$ shall be expected to be significant only for the V~3$d$ degrees of freedom.
Furthermore, considering the symmetry of our system, the $d_{z^2}$, $d_{x^2+y^2}+d_{xy}$ and $d_{xz}+d_{yz}$
components of the self-energy are not a-priori equal, as they belong to distinct irreducible representations of the  point symmetry group of the V atoms.
On the other hand, Eq.~\eqref{sloc} is a meaningful approximation provided that, for energies $\omega$ 
within $\sim$0.5~eV around the Fermi level,
the following hypothesis are verified:
\begin{itemize}
\item [1)] Most of the spectral weight arises from the V~3$d$ electrons.
\item [2)] The self-energy is approximately orbital-independent.
\item [3)] The momentum-independent local component $\Sigma_{loc}$ of the self-energy is approximately real and linear with respect to the frequency.
\end{itemize}

Here we use DFT and LDA+DMFT calculations to prove that these hypotheses are, in fact,  approximately applicable to our system.
We show the DFT calculated bands resolved with respect to their orbital character in Fig.~\ref{Supp:1}(a)-(c). These calculations indicate that the bands have mostly V~3$d$ character near the Fermi level. Specifically, the spectral weight is dominated by the $d_{z^2}$, $d_{x^2+y^2}+d_{xy}$ contributions.
Fig.~\ref{Supp:1}(d)  
illustrates the LDA+DMFT band structure obtained for a 
screened Hubbard interaction strength $U=6$~eV and a Hund's coupling constant $J=0.8$~eV. 
The orbitally-resolved LDA+DMFT local DOS in Fig.~\ref{Supp:1}(e) confirms that, consistent with DFT, most of the spectral weight near the Fermi level has V~3$d$ character.
Finally, as shown in Fig.~\ref{Supp:1}(f) and (g), the self-energy is approximately linear and similar for all of the V~3$d$ orbitals for energies $|\omega|\lesssim 0.5$~eV with respect to the Fermi level.
This observation is consistent with the LDA+DMFT quasi-particle weights:
\begin{equation}
Z_\alpha=\left|
1-\frac{\partial \Sigma'_{\alpha}}{\partial \omega}
\right|^{-1}
\,,
\end{equation}
see Table~1, which are all $\gtrsim 0.5$, in agreement with our simulation (in the range 0.52 to 0.54).

\begin{figure*}[t!]
\begin{center}
\includegraphics[width=.90\textwidth]{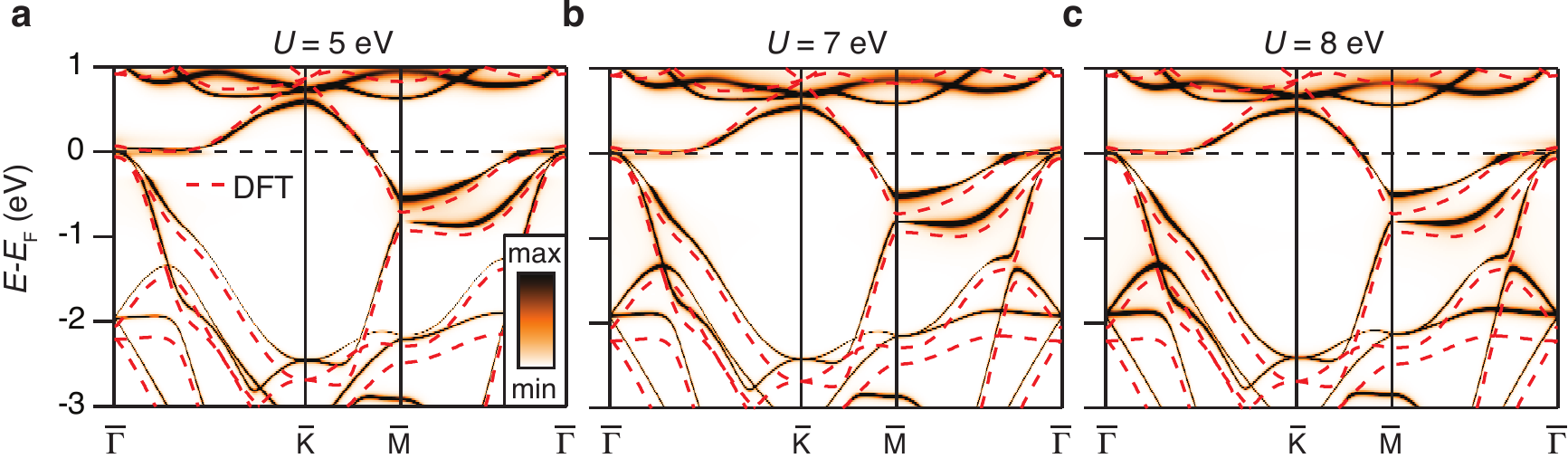}
\caption{\textbf{Behavior of LDA+DMFT bands as a function of $U$}. \textbf{a-c,} LDA+DMFT band structures calculated using the given values of screened on-site Coulomb interaction strength $U$, at $J=0.8$~eV. The bare DFT bands are also shown for comparison (red dashed lines).}
\label{Supp:2}
\end{center}
\end{figure*}

In Fig.~\ref{Supp:2} we show the LDA+DMFT bands for three different values of the Hubbard interaction strength $U$. The bands are found to be very similar for these $U$ values. This indicates that our theoretical predictions are robust.

\newpage
\subsection{Supplementary Section 2: Raw ARPES spectra in metallic and insulating phase}

\begin{figure*}[ht]
\begin{center}
\includegraphics[width=.9\textwidth]{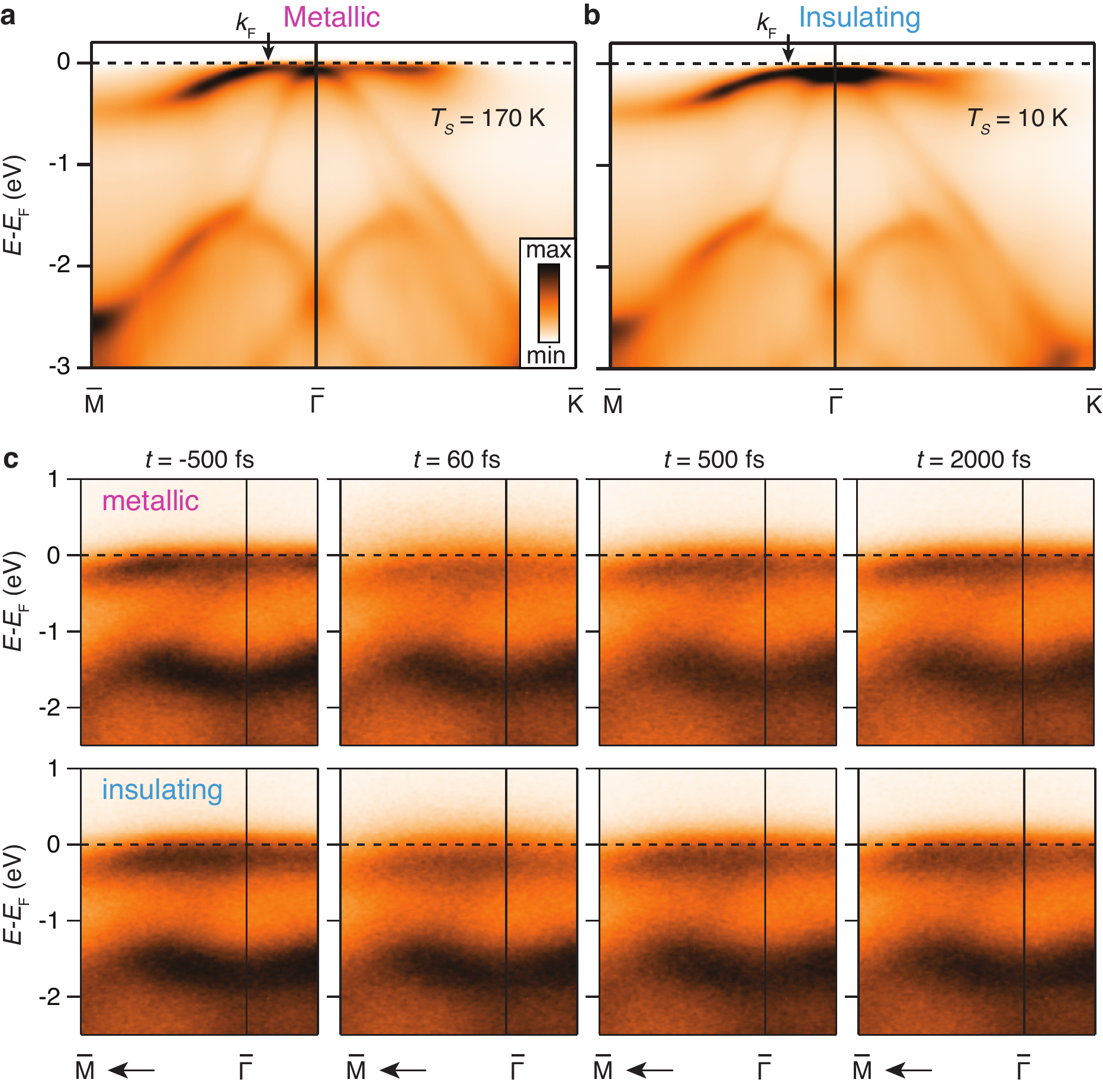}
\caption{\textbf{Comparison of ARPES spectra.} \textbf{a-b,} Static ARPES spectra for metallic and insulating phase.  \textbf{c,} TR-ARPES spectra at the given time delays for metallic (top) and insulating (bottom) phase.}
\label{Supp:3}
\end{center}
\end{figure*}

Fig. \ref{Supp:3}(a) and (b) show static ARPES spectra for both metallic and insulating phases. The dispersion of the top V 3$d$ band is different close to $k_{\mathrm F}$ due to the formation of the energy gap in the insulating phase. Figure \ref{Supp:3}(c) presents TR-ARPES snapshots of the spectral changes in these two scenarios before and after optical excitation.

\newpage
\subsection{Supplementary Section 3: Simulation of ARPES spectra}

\begin{figure*}[ht]
\begin{center}
\includegraphics[width=1\textwidth]{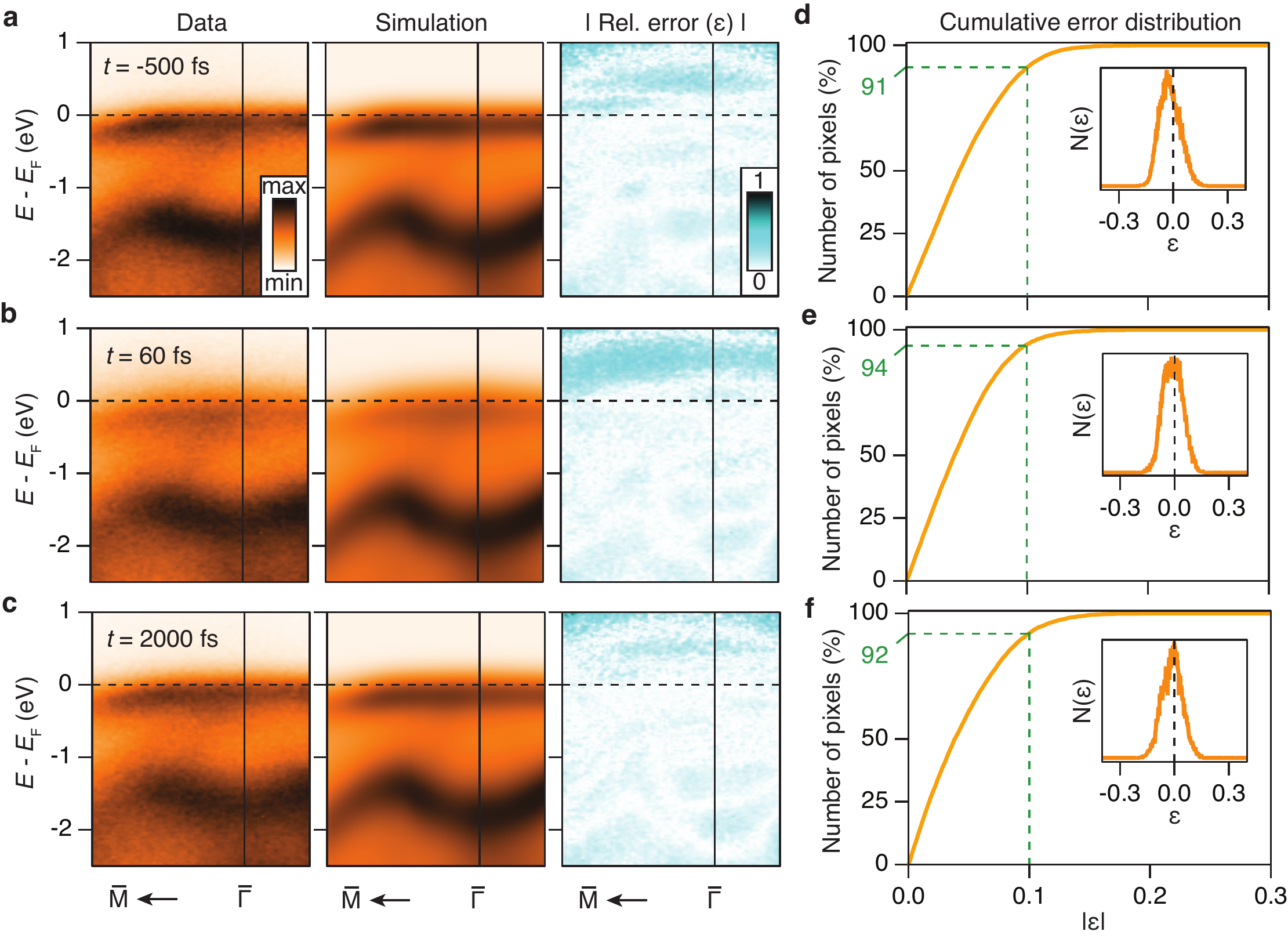}
\caption{\textbf{Quality of simulation.} \textbf{a-c,} TR-ARPES specta for $T_s=$ 200 K at the given time delays (left column). Simulated spectra and the corresponding unsigned relative errors ($\varepsilon=$ (simulation-data)/data) are shown in the middle and right columns, respectively.  \textbf{d}-\textbf{f,} Cumulative distribution of  $\mid\varepsilon\mid$ for the corresponding time-delays in the same row in \textbf{a}-\textbf{c} for the energy range -2.5 eV to \EF. The inserts present the distribution of $\varepsilon$. The green dashed lines in \textbf{d}-\textbf{f} correspond to $\mid\varepsilon\mid=0.1$.}
\label{Supp:4}
\end{center}
\end{figure*}

As mentioned in the methods section of the main text, the ARPES intensity can be expressed as,
\begin{equation}
{\cal I}_{ARPES}=[|{\cal M}_n(\vec{k}, \omega)|^2 {\cal A}_n(\vec{k}, \omega) n_{\mathrm FD}(\omega)]\ast {\cal R}_{\omega}\ast {\cal R}_k.
\label{eq:ARPES}
\end{equation}

\begin{figure*}[ht]
\begin{center}
\includegraphics[width=1\textwidth]{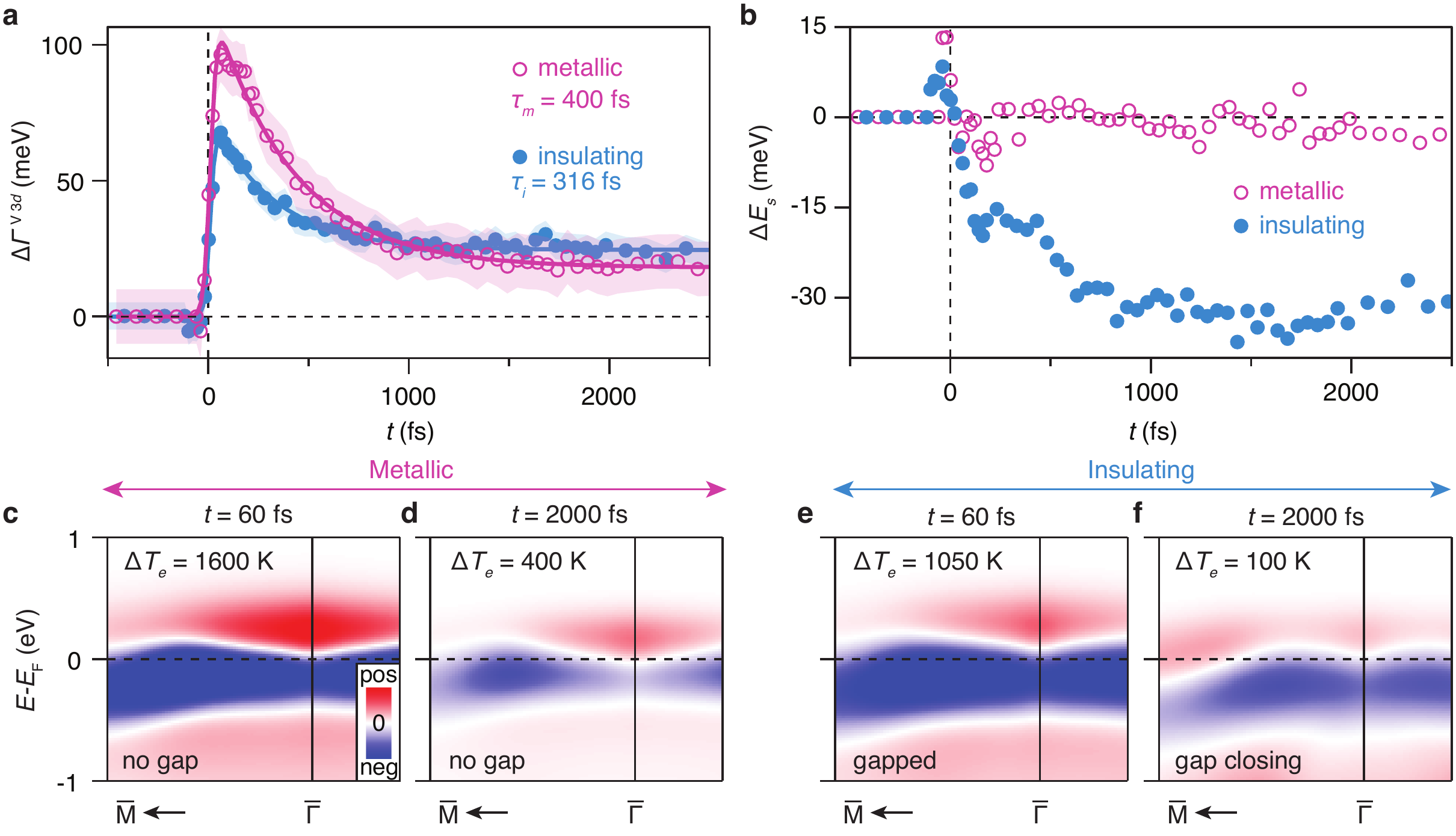}
\caption{\textbf{Resulting parameters from fits of \textit{t}-dependent photoemission intensity.} \textbf{a,}  The change in the scattering rate ($\Gamma$) of V 3$d$ states for both insulating ($T_s=$ 88 K) and metallic ($T_s=$ 200 K) phase. The solid curves show fits to a fast exponential rise followed by an exponential decay with the given time constants. \textbf{b,} The corresponding rigid energy shift of the spectra, ${\mathrm\Delta}E_s$. \textbf{c-d,} Difference between the fitted equilibrium spectra and the fitted spectra at the given time delays (60~fs and 2000~fs) for the metallic phase. \textbf{e-f,} Corresponding difference spectra for the insulating phase.}
\label{Supp:5}
\end{center}
\end{figure*}

The energy and momentum resolution functions (${\cal R}_{\omega}$ and ${\cal R}_k$) are known from instrument calibration and remain fixed for a given measurement. Furthermore, in static ARPES measurements we use that $T_e = T_s$ and $\mu =$ \EF~ such that $n_{FD}$ is fully specified. The parameters describing ${\cal M}_n(\vec{k},\omega)$ and $\Sigma_n(\vec{k},\omega)$ are obtained by performing a 2D fit of a simulated $(\omega,k)$-dependent intensity to the corresponding ARPES spectrum. Since the values of $Z$ and $\Delta$ at a given $T_s$ are intrinsic properties of the V $3d$ states that are independent of measurement configuration, we apply the values obtained from the static ARPES simulations to describe the TR-ARPES spectra. The parameters describing ${\cal M}_n(\vec{k},\omega)$ are related to the photoemission setup, however, we use the assumption that  ${\cal M}_n(\vec{k},\omega)$ is independent of time such that the matrix element is always determined in the equilibrium part of the TR-ARPES measurements. Data points acquired for $t\less-100$~fs are described using a single optimized spectrum, as the system is in equilibrium. The parameters of this optimized spectrum are used as input for the fit of the TR-ARPES data points acquired at the remaining time delay points.

In Figs. \ref{Supp:4}(a)-(c) we show the TR-ARPES spectra, simulated spectra and the corresponding unsigned relative error ($\mid\varepsilon\mid$) at $t=-500$~fs, 60~fs and 2000~fs for the metallic phase ($T_s=200$~K). The associated cumulative distribution of $\mid\varepsilon\mid$ is given in Figs. \ref{Supp:4}(d)-(f). As the actual intensity for the pixels above \EF~ is very small, irrespective of the simulation quality, the relative error  for these pixels are high. We have therefore selected the energy range from -2.5 eV to \EF~for our error analysis. All fitted pixels that fall below a margin set by $\mid\varepsilon\mid=0.1$ are deemed as providing a satisfactory agreement between model and data. We find that this is the case for $\approx$92\% of the pixels for all the three time delays. The symmetric distribution of the relative error (see inserts in Figs. \ref{Supp:4}(d)-(f)) with respect to $\varepsilon=0$ shows the unbiased nature of our simulation.

In Figs. \ref{Supp:5}(a) and (b) we show the changes in the remaining fit parameters - scattering rate ($\Gamma$) associated with the V 3$d$ band and the energy shift (${\mathrm\Delta}E_s$) which accompany the parameters $T_e$ and $\Delta$ shown in Fig. 3 of the main text. Figs. S5(c)-(f) present the intensity difference calculated by subtracting the fitted equilibrium spectra from the fitted spectra at $t=60$~fs and $t=2000$~fs, which may be compared with the experimental results in Figs. 2(e) and (f) of the main manuscript.

\newpage
\subsection{Supplementary Section 4: Reproducibility of intensity difference signals}

\begin{figure*}[ht]
\begin{center}
\includegraphics[width=1\textwidth]{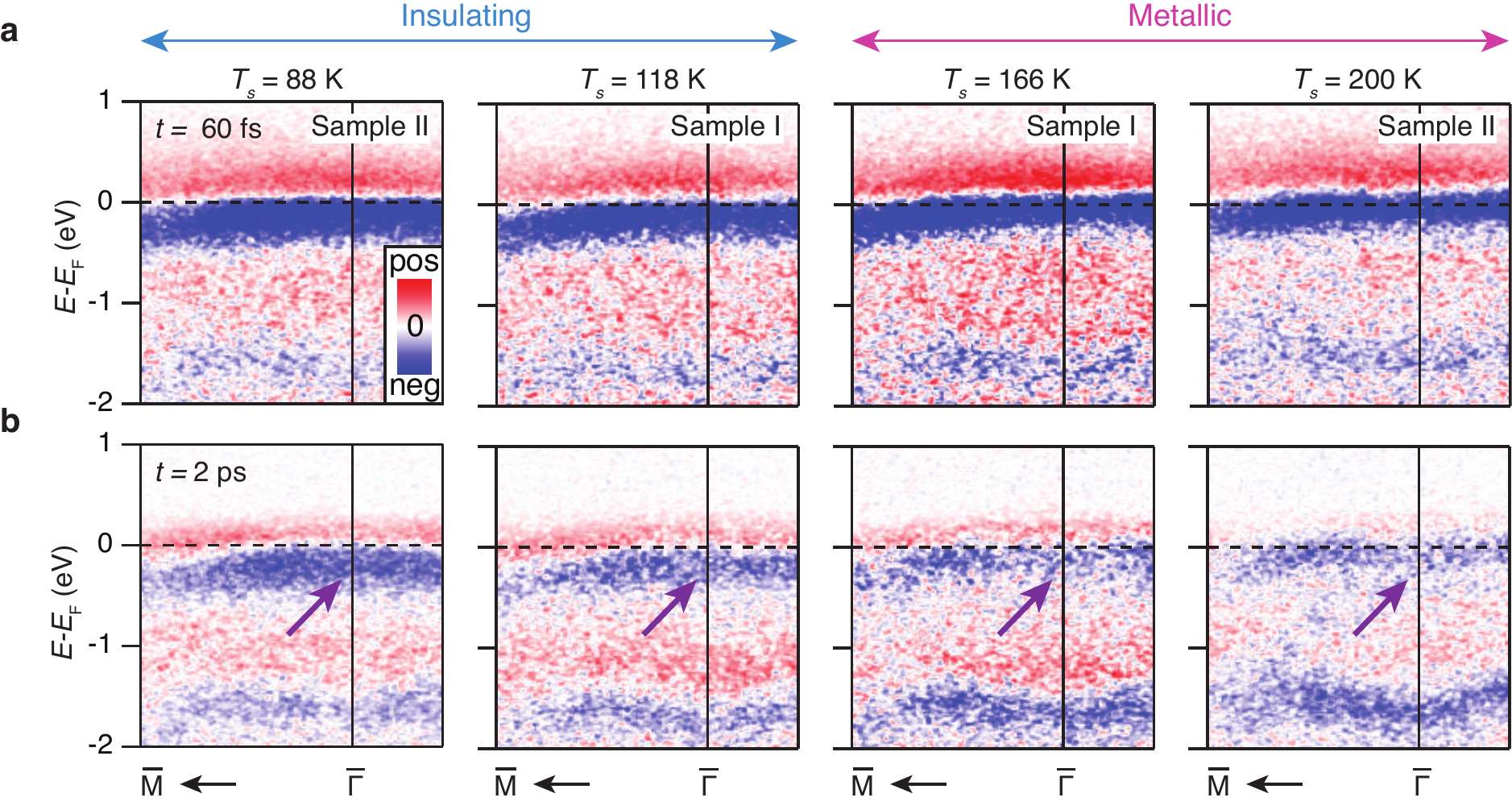}
\caption{\textbf{Reproducibility of the response to optical excitation.} \textbf{a,} Difference spectra at $t=60$ fs for sample temperature 88 K (sample II), 118 K (sample I), 166 K (sample I) and 200 K (sample II). \textbf{b,} Corresponding difference spectra at $t=2000$ fs. The purple arrows point to the $(\omega,k)$-region of the spectra most strongly affected by the phase transition.}
\label{Supp:6}
\end{center}
\end{figure*}

We have performed consistency checks of the observed intensity difference by repeating the measurements discussed in the main manuscript for sample temperatures $T_s$ of 88 K, 118 K, 166 K and 200 K and for two independent SL VSe$_2$ samples, verifying that the spectral signatures are robust for the two phases across $T_c=135$ K. Figure \ref{Supp:6} summarizes these results by presenting the corresponding intensity difference spectra.

\end{document}